\documentstyle[12pt,epsf]{article}
\begin{document}

\begin{titlepage}

\begin{flushright}
{\tt hep-th/0409004}\\
OCHA-PP-218\\
September, 2004
\end{flushright}

\vspace{5mm}

\begin{center}
{\Large \bf
Non-perturbative Aspects of Kaluza--Klein Modes 
\\in Five-dimensional Supersymmetric QCD on $S^1$
}

\vspace{2.5cm}
{\large Yukiko Ohtake}

\vspace{2.5mm}
{\it 
Department of Physics, Faculty of Science, Ochanomizu University\\
Institute of Humanities and Sciences, Ochanomizu University\\
2-1-1 Otsuka, Bunkyo-ku, Tokyo 112-8610, Japan}
\end{center}

\begin{abstract}
We study the stability of Kaluza--Klein (KK) modes in 
five-dimensional ${\cal N}=1$ supersymmetric QCD compactified on 
$S^1$ using the D3-brane probe realization.
We find a phenomenon in which 
the quark KK mode with the KK number $n=1$ decays 
while other KK modes are stable.
This is contrary to the ordinary assumption that 
the state with $n=1$ is the most stable in quark KK modes.
In addition, we show that a massive gauge singlet state
carrying the KK number exists stably.
This provides a proper candidate for dark matter.
\end{abstract}

\end{titlepage}

\section{Introduction}

Encouraged by great success of the brane-world scenario\cite{ADD},
many scientists have studied various models in which
our four-dimensional world is embedded in a higher dimensional spacetime.
In the research, the KK mode plays a central role since
it enables us to observe extra-dimensions from the four-dimensional
world.
Despite of the importance, however, the stability of KK modes
has been studied only perturbatively\cite{CMS}.
Thus we would like to investigate non-perturbative aspects of 
the stability in this paper.

For simplicity, we study five-dimensional $SU(2)$ Yang--Mills theory 
with $N_f$ fundamental matters compactified on $S^1$.
In addition we introduce ${\cal N}=1$ supersymmetry in five-dimension,
which provides ${\cal N}=2$ supersymmetry in the four-dimensional effective 
theory.
Then we analyze the model by using Seibreg--Witten theory\cite{SW},
which tells us the exact effective theory and the stability of the states 
in it.
Although determining the stability is complicated in field theory,
we can easily determine it in IIB string theory, where
the effective theory appears on the D3-brane in 
a 7-brane background\cite{BFMNS}.

In the following we review how the theory appears on the D3-brane 
and how the stability is determined.
After that, we apply the method to study the stability of phenomenologically
important KK states. 
In section 3, we determine the stability of quark KK modes.
As a result, we find a phenomenon in which the KK mode with the
KK number $n=1$ decays while other KK modes are stable.
In section 4, we show that there exists a massive gauge singlet,
which gives a proper candidate for dark matter.

\section{Brane Construction of Five-dimensional SQCD}
Five-dimensional ${\cal N}=1$ $SU(2)$ QCD appears in IIA string theory 
as the world volume theory of a D4-brane in the 8-brane background
consisting of an O8-plane and $N_f$ D8-branes parallel to the 
D4-brane\cite{Seiberg}.
In this picture, a string connecting a D8-brane and the D4-brane 
provides a quark, while a string with the both ends on the D4-brane
provides a gauge field.
As we set the origin of the space transverse to the 8-branes 
at the position of the O8-plane,
the positions of D8-branes correspond to the quark bare mass parameters 
and that of D4-brane corresponds to the moduli parameter which comes from 
the vacuum expectation value of an adjoint Higgs field.

To compactify the field theory on $S^1$, we compactify the IIA string theory
on $S^1$ parallel to the D4-brane.
Moreover we perform T-duality along $S^1$ to obtain the string picture 
which describes the effective four-dimensional theory well.
The T-duality transforms IIA theory into IIB theory,
the $N_f$ D8-brane into $N_f$ D7-branes, the O8-plane into two O7-planes, 
and the D4-brane into a D3-brane, on which the effective theory appears.
As the five-dimensional case, 
the positions of branes correspond to the parameters of the field theory 
and strings ending on the D3-brane provide quarks and gauge fields.
In addition,
the distance between the O7-planes is related to $1/R$ where 
$R$ is the radius of $S^1$,
and the winding number of strings around the O7-planes corresponds
to the KK number in the field theory.

Moreover, there appear various strings providing solitons 
on the D3-brane\cite{Schwarz}.
To see this, we introduce the notation used in \cite{GZ}:
we call a fundamental string $(1,0)$, a D7-brane $[1,0]$,
the $SL(2,Z)$ dual strings $(p,q)$ and the dual 7-branes $[p,q]$, 
where $p$ and $q$ are relatively prime integers.
As $(1,0)$ emanates from $[1,0]$, $(p,q)$ emanates from $[p,q]$.
In this notation the background of $N_f$ D7-branes and two O7-planes
is written as $[1,0]^{N_f}([2,-1][0,1])^2$
because an O7-plane is identified with $[2,-1][0,1]$\cite{Sen}.
Thus there appear not only $(1,0)$ but also $(2,-1)$, $(0,1)$ 
and their composites.
One of these strings, $(p,q)$, is observed as a particle with the
electric charge of unbroken $U(1)$ gauge symmetry of $SU(2)$ $p$
and the magnetic charge $q$.
What composites exist stably depends on details of the 7-brane
background.
In the following we shall review a gravitational description 
of the background.

\subsection{7-brane backgrounds}

A 7-brane background is expressed as a solution of the ten-dimensional
supergravity which is the low energy effective theory of IIB 
theory\cite{cosmic}.
The solution consists with an eight-dimensional flat spacetime
parallel to the 7-branes and a two-dimensional curved space denoted 
by $z$-plane.
The $z$-plane is described by an elliptic curve,
\begin{equation}
y^2= x^3 +f(z)x +g(z),
\label{eqn:elliptic}
\end{equation}
where $x$ and $y$ are complex parameters, and $f(z)$ and $g(z)$ are
polynomials of $z$.
The equation (\ref{eqn:elliptic}) gives a torus for a fixed value of $z$.
The periods of the torus are
\begin{equation}
\frac{da(z)}{dz}=\oint_{\alpha} \frac{dx}{y}, \hspace{5mm}
\frac{da_D(z)}{dz}= \oint_{\beta} \frac{dx}{y},
\end{equation}
where $\{\alpha, \beta\}$ is an orthogonal basis of two-cycles
of the torus.
Using $a(z)$ and $a_D(z)$, we obtain the vacuum expectation value 
of the dilaton $\phi$ and the Ramond-Ramond zero-form $\chi(z)$,
and the metric on the $z$-plane $ds^2$ as follows;
\begin{equation}
\tau(z)=ie^{-\phi(z)}¡Ü\chi(z)=\frac{da_D(z)}{da(z)},
\label{eqn:tau}
\end{equation}
\begin{equation}
ds^2= \mbox{Im}\tau(z)\left|\frac{da(z)}{dz}dz\right|^2.
\label{eqn:metric}
\end{equation}

The positions of 7-branes are zeros of the discriminant of 
(\ref{eqn:elliptic}), $\Delta(z)=4f(z)^3+27g(z)^2$.
The number of 7-branes at a zero is the order of the zero.
Types of 7-branes at a zero of $\Delta(z)$ are 
determined from the monodromy matrix acting on 
$^t(\frac{da_D}{dz}, \frac{da}{dz})$
when we go around the zero anti-clockwisely\cite{GZ}.
When the matrix at an $n$-th order zero is 
expressed as $M_{[p_1, q_1]}M_{[p_2, q_2]}\cdots M_{[p_n, q_n]}$,
where
\begin{equation}
M_{[p,q]}=\left(
\begin{array}{cc}
1-pq & p^2\\
-q^2 & 1+pq
\end{array} 
\right),
\label{eqn:mono-f}
\end{equation}
we decide that the 7-branes at the zero are
$[p_1, q_1][p_2, q_2]\cdots [p_n, q_n]$.

\subsection{string junctions}

In this subsection we introduce strings in a 7-brane background.
If a string is stable, it stretches along a certain curve on the $z$-plane 
to minimize its mass.
We call such curve geodesic in this paper.
If a geodesic passes $z=u$, the corresponding string can end on the D3-brane
at $z=u$
and provides a stable particle in the world volume theory with the 
moduli parameter $u$.
Therefore the region on the $z$-plane where the geodesic of a 
string $(p,q)$ passes 
corresponds to the region on the moduli $u$-plane where the particle
$(p,q)$ exists stably.
Hence we determine geodesics in (\ref{eqn:metric}) to investigate 
the stability of particles. 

We start with a simplest background, a 7-brane $[p,q]$ at $z=z_1$.
In this case, the string emanating from the background is $(p,q)$.
The mass of the string connecting the 7-brane and the D3-brane at 
$z=u$ along a curve $C$ is given by 
\begin{equation}
M_{(p,q)}(u)=\int_C T_{p,q}ds,
\label{eqn:mass-f}
\end{equation}
where $T_{p,q}=|p-q\tau|/\sqrt{\mbox{Im}\tau}$ is 
the tension of $(p,q)$\cite{Schwarz}.
Substituting (\ref{eqn:tau}) and (\ref{eqn:metric}) for (\ref{eqn:mass-f}),
we obtain the lower bound of the mass as follows:
\begin{eqnarray}
M_{(p,q)}(u)
& =&\int_C \left|p\frac{da(z)}{dz}-q\frac{da_D(z)}{dz}\right|\left|dz\right|
\nonumber \\
&\geq& \left|pa(u)-qa_D(u)-pa(z_1)+qa_D(z_1)\right|.
\label{eqn:mass} 
\end{eqnarray}
The minimum is realized if and only if all the points on $C$ satisfy
\begin{equation}
\mbox{Arg}\left(pa(z)-qa_D(z)-pa(z_1)+qa_D(z_1)\right)=\phi,
\label{eqn:config}
\end{equation}
where $\phi$ is a constant between 0 and $2\pi$.

The equation (\ref{eqn:config}) for a fixed value of $\phi$ gives
a geodesic of $(p,q)$.
In this background we can use $pa-qa_D\sim c(z-z_1)$\cite{SW},
so that (\ref{eqn:config}) gives a half line from $z=z_1$ 
at an angle of $\phi-\mbox{Arg}c$ to the positive real axis.
Therefore, as $\phi$ varies from 0 to $2\pi$,
the geodesic sweeps all the $z$-plane.
This means that the state corresponding to the string $(p,q)$
appears stably in the theory with any value of the moduli parameter $u$.

Next we consider the background of two 7-branes, $[p_1,q_1]$ at $z=z_1$
and $[p_2,q_2]$ at $z=z_2$.
We assume $z_1\neq z_2$ because two 7-branes never coincide unless
$[p_1,q_1][p_2,q_2]=[1,0][1,1]$, $[1,0][1,0]$ or their $SL(2,Z)$ 
dual\cite{Kodaira}, and 
geodesics for the exceptions are derived from the limit 
$z_1\rightarrow z_2$ in the case of $z_1\neq z_2$.
In this background there exists a string $(p_i,q_i)$ ($i=1,2$) 
connecting $[p_i,q_i]$ at $z=z_i$ and the D3-brane at $z=u$.
The geodesic is described by (\ref{eqn:config}) where we replace
$(p,q)$ with $(p_i,q_i)$ and $z_1$ with $z_i$.
The equation is difficult to solve, however, 
qualitatively the geodesic should be a line bent by gravity 
of the other 7-brane.
Thus the geodesic sweeps all the $z$-plane as in the background of 
a 7-brane.
Then the particle $(p_i,q_i)$ appears stably in the theory 
with any value of the moduli parameter.

In addition to the strings, there exists a string junction\cite{Schwarz2}.
A string junction is constructed from $n_1(p_1,q_1)$ and 
$n_2(p_2,q_2)$ emanating from the 7-branes and merge into a
string $(p,q)=\sum_{i=1,2}n_i(p_i,q_i)$ as in Fig.1a.
\begin{figure}
\hspace{3cm}
\epsfxsize=8.4cm
\epsfbox{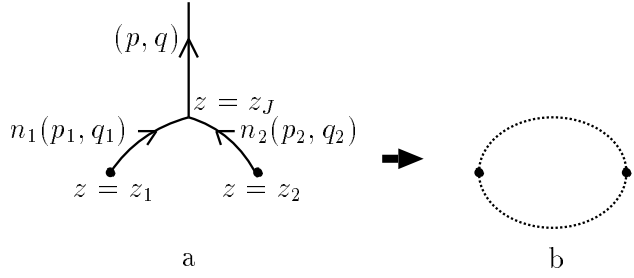}
\caption{A configuration of string junction (a) and the related CMS (b).}
\end{figure}
The mass of a junction is the sum of the component string masses.
These are derived similar to (\ref{eqn:mass}) 
and the geodesics are descrived by 
\begin{eqnarray}
\mbox{Arg}\left(p_ia(z)-q_ia_D(z)-p_ia(z_i)+q_ia_D(z_i)\right)&=&\phi_i, 
\label{eqn:config2-1}\\
\mbox{Arg}\left(pa(z)-qa_D(z)-pa(z_J)+qa_D(z_J)\right)&=&\phi,
\label{eqn:config2-2}
\end{eqnarray}
where $z=z_J$ is the point where the strings merge.
Thus the mass of the string junction is given by 
\begin{eqnarray}
M_{(p,q)}(u)&=&|pa(u)-qa_D(u)-pa(z_J)+qa_D(z_J)|\nonumber\\
&&+\sum_{i=1,2}n_i|p_ia(z_J)-q_ia_D(z_J)-p_ia(z_i)+q_ia_D(z_i)|\nonumber\\
&\geq& \left|pa(u)-qa_D(u)-\sum_{i=1,2}n_i(p_ia(z_i)-q_ia_D(z_i))\right|,
\label{eqn:mass2}
\end{eqnarray}
where the lower bound is realized when $\phi_1=\phi_2=\phi$.
Therefore the geodesics (\ref{eqn:config2-1}) and (\ref{eqn:config2-2}) are 
rewritten as follows;
\begin{eqnarray}
\mbox{Arg}\left(p_ia(z)-q_ia_D(z)-p_ia(z_i)+q_ia_D(z_i)\right)&=&\phi,
\label{eqn:config2-3} \\
\mbox{Arg}\left(pa(z)-qa_D(z)-\sum_{i=1,2}n_i(p_ia(z_i)-q_ia_D(z_i))
\right)&=&\phi.
\label{eqn:config2-4}
\end{eqnarray}

The geodesics (\ref{eqn:config2-3}) and (\ref{eqn:config2-4}) 
for a proper value of $\phi$ 
are estimated as in Fig.1a.
If we change $\phi$ continuously, the geodesic of $(p_i,q_i)$ rotates
around $z=z_i$.
Then the merging point $z=z_J$ draws a circle as depicted in Fig.1b,
and the string $(p,q)$ sweeps only the outside of the circle.
Hence the particle $(p,q)$ exists when the moduli parameter $u$
takes a value outside the circle and disappears when $u$ takes
a value inside it.
When $u$ takes a value on the circle, 
$(p,q)$ becomes marginally stable:
in this case, the D3-brane is located at $z=z_J$ in Fig.1a,
so that $(p,q)$ can decay to $n_1$ $(p_1,q_1)$ and $n_2$ $(p_2,q_2)$.
Then the curve drawn by $z=z_J$ is called a curve of marginal stability
(CMS).
Since a point on CMS satisfies (\ref{eqn:config2-3}) and (\ref{eqn:config2-4}),
CMS obeys
\begin{equation}
\mbox{Im}\frac{p_ia(z)-q_ia_D(z)-p_ia(z_i)+q_ia_D(z_i)}{pa(z)-qa_D(z)-\sum_{j=1,2}n_j(p_ja(z_j)-q_ja_D(z_j))}=0.
\end{equation}

Finally, we consider a background of $N$ 7-branes, $[p_i,q_i]$ at $z=z_i$
($i=1,2,\ldots,N(\geq 3)$).
As discussed before, a string emanating from each 7-branes, $(p_i,q_i)$, 
sweeps all the $z$-plane. 
Then the corresponding state exists in the field theory with any value
of $u$.
Moreover, there exist string junctions.
We call $(p,q)_{\vec{n}}$ a string junction constructed from $n_i(p_i,q_i)$,
where $(p,q)=\sum_{i=1}^{N}n_i(p_i,q_i)$ and $\vec{n}=(n_1,n_2,\ldots,n_N)$.
The configuration is complicated because $(p_i, q_i)$ merge one after another.
To fix the order in which the strings merge, 
we have to solve quantitatively the geodesics of $(p_i,q_i)$
which are described by (\ref{eqn:config2-3}).
On the other hand, the mass and the geodesics of the component
strings are the same form independent of the order;
\begin{equation}
M_{(p,q)_{\vec{n}}}(u)=\left|pa(u)-qa_D(u)-\sum_{i=1}^{n}n_i(p_ia(z_i)-q_ia_D(z_i))\right|,
\label{eqn:mass2-4}
\end{equation}
\begin{equation}
\mbox{Arg}\left(p'a(z)-q'a_D(z)-\sum_{i=1}^{N}n'_i(p_ia(z_i)-q_ia_D(z_i))
\right)=\phi,
\label{eqn:config2-5}
\end{equation}
where $(p',q')_{\vec{n'}}$ runs all the component strings.
From (\ref{eqn:config2-5}) we see that CMS is given by 
\begin{equation}
\mbox{Im}\frac{p'a(z)-q'a_D(z)-\sum_{j=1}^{N}n'_j(p_ja(z_j)-q_ja_D(z_j))}{pa(z)-qa_D(z)-\sum_{j=1}^{N}n_j(p_ja(z_j)-q_ja_D(z_j))}=0,
\end{equation}
where $(p',q')_{\vec{n'}}$ is a string merging the outgoing string 
$(p,q)_{\vec{n}}$.
To specify $(p',q')_{\vec{n'}}$, we have to fix the order in which strings 
merge. 
In the following section, we do this in some backgrounds.

\section{Stability of quark Kaluza--Klein modes}

In the previous section, we review a method of determining the stability of 
particles appearing on the D3-brane in a 7-brane background.
Using the method, we shall derive the spectrum of quark KK modes in 
five-dimensional ${\cal N}=1$ $SU(2)$ QCD compactified on $S^1$, which 
is realized on the D3-brane in the 7-brane background 
$[1,0]^{N_f}([2,-1][0,1])^2$.
For simplicity, we concentrate on the cases in which $N_f=7,6,5$, all 
the quarks are massless, and the gauge coupling constant $g\rightarrow\infty$.
In these cases the global symmetry is enhanced to $E_{N_f+1}$,
and all the 7-branes but one coincide to produce the 
symmetry\cite{fg}.

The explicit forms of $a$ and $a_D$ for the backgrounds are derived
in \cite{MOY} as follows; 
\begin{equation}
\left(
\begin{array}{c}
a_D(z)\\ a(z)
\end{array}
\right)=
\frac{\sin\pi\alpha}{2\pi(1-2\alpha)R}
\left(
\begin{array}{cc}
2\xi_2& -2\xi_1\\
-\frac{\omega}{\sin\pi\alpha}\xi_2& \frac{\bar{\omega}}{\sin\pi\alpha}\xi_1
\end{array}
\right)
\left(
\begin{array}{c}
V_1(z)\\V_2(z)
\end{array}
\right),
\label{eqn:SWperiods}
\end{equation}
where $\xi_1=\Gamma(2\alpha)/\Gamma^2(\alpha)$,
$\xi_2=\Gamma(2-2\alpha)/\Gamma^2(1-\alpha)$,
$\omega=e^{i\pi\left(\frac{1}{2}-\alpha\right)}$ and $\alpha=1/6,1/4,1/3$ 
for $N_f=7,6,5$, respectively.
Also $V_i(z)$($i=1,2$) are given by 
\begin{eqnarray}
V_1(z)&=&\frac{1}{\alpha}z^{\alpha}{}_3F_2(\alpha, \alpha, \alpha;2\alpha, 1+\alpha;z)\\
V_2(z)&=&\frac{1}{1-\alpha}z^{1-\alpha}{}_3F_2(1-\alpha, 1-\alpha, 1-\alpha;
2(1-\alpha), 2-\alpha; z).
\end{eqnarray}
In these backgrounds 7-branes are located at $z=0$ and $z=1$
\footnote{The distance between the 7-branes actually depends on the radius
$R$.
Here we eliminate it by rescaling the original $z$ to 
$\frac{2}{27}R^6z$, $\frac{1}{8}R^4z$, $\frac{4i}{27}R^3z$ 
for $N_f=7,6,5$, respectively.}.
The monodromy matrices around these points are 
\begin{equation}
M_0=\left(
\begin{array}{cc}
N_f-7& N_f-8\\
1&1
\end{array}
\right),\hspace{5mm}
M_1=\left(
\begin{array}{cc}
1& 0\\
-1&1
\end{array}
\right).
\label{eqn:matrix}
\end{equation}
These are rewritten as
$M_0=M_{[1,0]}^{N_f}M_{[2,-1]}M_{[0,1]}M_{[2,-1]}$ and 
$M_1=M_{[0,1]}$, where $M_{[p,q]}$ is given in (\ref{eqn:mono-f}).
Thus 7-branes $[1,0]^{N_f}[2,-1][0,1][2,-1]$ are located at $z=0$
and a 7-brane $[0,1]$ at $z=1$. 
For the later use we also provide the monodromy matrices acting on 
$^t(a_D, a, 1/R)$; 
\begin{equation}
\tilde{M}_0=\left(
\begin{array}{ccc}
N_f-7&N_f-8&0\\
1&1&0\\
0&0&1
\end{array}
\right),\hspace{5mm}
\tilde{M}_1=\left(
\begin{array}{ccc}
1&0&0\\
-1&1&1\\
0&0&1
\end{array}
\right).
\label{eqn:mono-t}
\end{equation}

In this background there exists a string junction constructed from $(p,q-n)$
emanating from $z=0$ and $n(0,1)$ from $z=1$.
The mass formula is obtained from (\ref{eqn:mass2-4}),
\begin{eqnarray}
M_{(p,q)_n}&=&
|pa(u)-qa_D(u)-pa(0)+(q-n)a_D(0)+na_D(1)|\nonumber\\
&=&\left|pa(u)-qa_D(u)+\frac{n}{R}\right|,
\label{eqn:mass4}
\end{eqnarray}
where we have used $a(0)=a_D(0)=0$ and $a_D(1)=1/R$.
The geodesics of the outgoing string of the junction 
is derived from (\ref{eqn:config2-5}), 
\begin{equation}
\mbox{Arg}\left(pa(z)-qa_D(z)+\frac{n}{R}\right)=\phi.
\label{eqn:config3}
\end{equation}
We call $(p,q)_n$ a string whose configuration satisfying (\ref{eqn:config3})
throughout this paper.
In this notation a string $(1,0)$ from $z=0$ which corresponds to 
a quark is represented by $(1,0)_0$.
In addition, a string corresponding to a quark KK mode with 
the KK number $n$ is $(1,0)_n$,
since the number of strings from $z=1$ is identified with the KK number 
from (\ref{eqn:mass4}).

Here we introduce branch cuts on the $z$-plane to make the functions
$a(z)$ and $a_D(z)$ single-valued\cite{GHZ}.
We take the cuts running from each 7-brane to $z=\infty$ along the 
real axis.
If we go across the cut emanating from $z=i$ ($i=0,1$) from the 
lower half-plane, $^t(a_D, a, 1/R)$ changes to 
$\tilde{M}_{i}^{-1}{}^t(a_D, a, 1/R)$.
Simultaneously, we should change $^t(p,q,n)$ into $K_i{}^t(p,q,n)$ where
\begin{equation}
K_0=\left(
\begin{array}{ccc}
1&8-N_f&0\\
-1&N_f-7&0\\
0&0&1
\end{array}
\right), \hspace{5mm}
K_1=\left(
\begin{array}{ccc}
1&0&0\\
1&1&0\\
1&0&1
\end{array}
\right),
\label{eqn:mono2}
\end{equation}
to keep the geodesic (\ref{eqn:config3}) unchanged.
We assume in the following 
that the cut from $z=0$ is infinitesimally lower than the cut from $z=1$.
Thus $^t(p,q,n)$ changes to $K_1K_0{}^t(p,q,n)$ when we go across the 
cuts on $\mbox{Re}z>1$ from the lower half-plane.

With these branch cuts, we can determine $a$ and $a_D$ numerically 
using Mathematica and solve the geodesics (\ref{eqn:config3}) approximately.
To do this, we take a lattice of the spacing $d$ on the $z$-plane for the
region $|\mbox{Re}z|\leq s$ and $|\mbox{Im}z|\leq s$. 
Then we generate data of the numerical values of $a$ and $a_D$ at the lattice
points, and make data of signs of $\mbox{Im}(pa-qa_D-n)/e^{i\phi}$.
If the signs at both ends of a link are different, at least a solution of 
(\ref{eqn:config3}) exists on the link.
Thus we look for such links and plot the middle of each as an approximate 
solution.
These plots tell us an approximate configuration of $(p,q)_n$ at $\phi$.
We made plots for $d=0.01$ and $s=5$ for $N_f=7,6,5$ independently,
and found that the  geodesics in $N_f=7,6,5$ are similar.
Then we shall give a figure to show the results of the three cases at once.

Now we are ready to consider our main problem, the configuration of 
$(1,0)_n$.
First we derive the geodesics (\ref{eqn:config3}) for $(1,0)_0$.
Taking an appropriate value of $\phi$, we find that the configuration of
$(1,0)_0$ is as depicted in Fig.2a.
The geodesic crosses the cuts on $\mbox{Re}z>1$, then the set of charges 
$^t(1,0,0)$ changes to $K_1K_0{}^t(1,0,0)=(1,0,1)$.
Thus we join $(1,0)_0$ to $(1,0)_1$ on the cuts.
When we change $\phi$ gradually,
we find that the geodesic $(1,0)_0$ sweeps the whole $z$-plane.
Therefore the quark $(1,0)_0$ exists in the theory with any value
of the moduli parameter.
\begin{figure}
\hspace{0.5cm}
\epsfxsize=15cm
\epsfbox{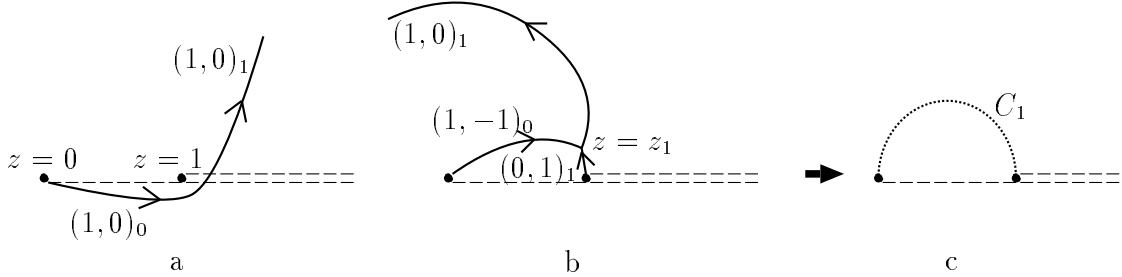}
\caption{$(1,0)_1$ and the related CMS.}
\end{figure}

Second we consider how $(1,0)_1$ appears.
Recall that it already appears in Fig.2a as $(1,0)_0$ crosses
the cuts on $\mbox{Re}z>1$.
Therefore, as we change $\phi$ in Fig.2a, $(1,0)_1$ of the string 
sweeps the right-hand side of the upper-half $z$-plane until the 
string collides with the 7-brane at $z=1$.
If we change $\phi$ further after the collision, the string changes 
to a string junction with the outgoing string $(1,0)_1$ as depicted in 
Fig.2b\cite{GZ,BF2}.
This is understood in the following way.
After the collision, $(1,0)_0$ gets over the cut on $\mbox{Re}z<1$
and the set of charges $(1,0,0)$ changes to $K_0{}^t(1,0,0)=(1,-1,0)$.
Thus $(1,0)_0$ changes to $(1,-1)_0$ as depicted in Fig.2b
while $(1,0)_1$ preserves its charges.
Consequently, to fill the difference of these charges,
$(0,1)_1$ should be created at $z=1$.
As we change $\phi$ in Fig.2b, a point where strings merge,
say $z=z_1$, draws a half circle $C_1$ as in Fig.2c.
Simultaneously $(1,0)_1$ sweeps all the $z$-plane except for the 
region $S_1$ surrounded by $C_1$ and the real axis.
All the lower-half plane is swept because $(1,0)_1$ is hardly bent 
by 7-branes at $z=0$, which act as a large gravitational source.
From these facts, we conclude that the quark KK mode $(1,0)_1$
exists outside $S_1$, decays to $(1,-1)_0+(0,1)_1$ on $C_1$,
and disappears inside $S_1$.

Third we consider $(1,0)_2$.
The string appears in Fig.3a as $(1,0)_1$ crosses the 
cuts on $\mbox{Re}z>1$.
As we change $\phi$ in Fig.3a, 
the outgoing string collides the 7-brane at $z=1$ 
and changes to another string junction as depicted in Fig.3b.
During the transition, an additional $(0,1)_1$ is created.
The geodesics of two $(0,1)_1$ are the same, 
but we separate them in Fig.3b to show the structure of the junction.
One of $(0,1)_1$ merges with $(1,-1)_1$ which appears as $(1,0)_1$ crosses
the cut while the other merges with $(1,-1)_0$.
Changing $\phi$ further, we find that the merging point of $(1,-1)_1$ and
$(0,1)_1$, say $z=z_2$, draws a half circle $C_2$ as depicted in Fig.3c.
Also we find that the outgoing string $(1,0)_2$ sweeps all the $z$-plane
except for the region $S_2$ surrounded by $C_2$ and the real axis.
Thus we conclude that $(1,0)_2$ exists outside $S_2$,
decays to $(1,-1)_1+(0,1)_1$ on $C_2$, and disappear inside $S_2$.
\begin{figure}
\hspace{0.5cm}
\epsfxsize=15cm
\epsfbox{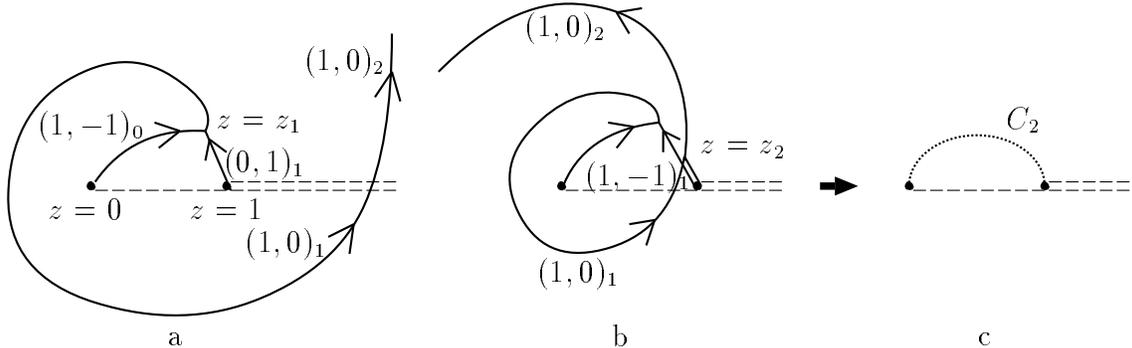}
\caption{$(1,0)_2$ and the related CMS.}
\end{figure}

Repeating the operation, we can construct $(1,0)_n$ for $n\geq 3$. 
The results are similar to the cases of $n=1,2$.
A geodesic $(1,0)_n$ appears as $(1,0)_{n-1}$ crosses the cuts on 
$\mbox{Re}z>1$.
As we change $\phi$, the outgoing string becomes a junction in 
which $(0,1)_1$ and a string junction $(1,-1)_{n-1}$ merge into $(1,0)_n$
at a point.
The merging point draws a half circle $C_n$ connecting $z=0$ and $z=1$.
Simultaneously, $(1,0)_n$ sweeps all the $z$-plane except for the
region $S_n$ surrounded by $C_n$ and the real axis.
Thus the quark KK mode $(1,0)_n$ exists outside $S_n$,
decays to $(1,-1)_{n-1}+(0,1)_1$ on $C_n$,
and disappears inside $S_n$.
We checked the result for $n=3,4$ numerically and
for other $n$ by using the approximation introduced in \cite{GHZ},
where we neglect the effects of the 7-brane at $z=1$ on the 
metric.

Here we compare the sizes of $C_1$ and $C_2$.
For this purpose, let us notice the geodesic $(0,1)_1$ in Fig.3b
and two points on it, $z=z_1$ and $z=z_2$.
If we change $\phi$, $(0,1)_1$ rotates around $z=1$ and each point 
$z=z_i$ ($i=1,2$) draws a half circle $C_i$.
We see that $C_1$ is larger than $C_2$ 
since $z=z_1$ is farther from $z=1$ than $z=z_2$. 
Similar consideration is valid for other pairs of positive KK numbers, 
$n_1$ and $n_2$.
Then we conclude that $C_{n_1}$ is larger than $C_{n_2}$ when $n_1<n_2$
(Fig.4).
Accordingly, $(1,0)_1$ disappears first and other states
$(1,0)_n$ ($n\geq 2$) disappear in numeric order of $n$
when we change the moduli parameter from a value above $C_1$ to 
the segment $[0,1]$ on the real axis.
This result is contrary to the ordinary assumption that $(1,0)_1$ is
the most stable in quark KK modes.
\begin{figure}
\hspace{5.5cm}
\epsfxsize=5cm
\epsfbox{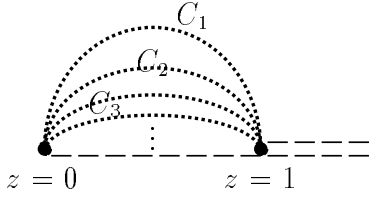}
\caption{CMS of quark KK modes.}
\end{figure}

Before concluding this section,
we comment on the quark KK modes with the negative KK numbers,
$(1,0)_{-n}$ ($n\geq 1$).
The outgoing string $(1,0)_{-n}$ 
appears when we rotate $(1,0)_0$ clockwisely $n$ times.
The resulting configuration of the junction is the mirror image of the
configuration of $(1,0)_n$ with respect to the real axis.
Actually, the replacement $z\rightarrow \overline{z}$ in (\ref{eqn:config3})
corresponds to $(p,q)_n\rightarrow (p+(8-N_f)q, -q)_{-n}$\cite{Ohtake}.
Thus $(1,0)_{-n}$ exists outside the mirror image of $S_n$,
decays to $(N_f-7,1)_{-n+1}+(8-N_f,-1)_{-1}$ on the mirror image of $C_n$,
and disappears inside the mirror image of $S_n$.

\section{Stability of a gauge singlet state}

In this section we study a loop string encircling all the 7-branes.
The string has a finite length so that it should have finite mass.
In addition, when we locate the D3-brane on the loop, 
the total string charges absorbed by the brane are apparently zero.
Then the string corresponds to a massive gauge singlet in field theory, 
and provides a proper candidate for dark matter.
This motivate us to study the string.

First we consider a loop string in the simple backgrounds given by 
(\ref{eqn:SWperiods}).
The mass of a string stretching along a loop $C$ surrounding all the 
7-branes is derived as follows;
\begin{eqnarray}
M(u)&=&\oint_C T_{1,0}ds \nonumber\\
&\geq&\lim_{\epsilon\rightarrow 0+}
\left| a(z_*-i\epsilon)-a(z_*+i\epsilon)\right|\nonumber\\
&=& \lim_{\epsilon\rightarrow 0+}
\left| a(z_*+i\epsilon)+\frac{1}{R}-a(z_*+i\epsilon)\right|=\frac{1}{R},
\end{eqnarray}
where $z=z_*$ is the intersection of $C$ and the real axis on 
$\mbox{Re}z\geq 1$,
and we have used (\ref{eqn:mono-t}) to derive the third line.
The lower bound is realized when all the points on $C$ satisfy
\begin{equation}
\mbox{Arg}\left(a(z)-a(z_*+i\epsilon)\right) =0,
\label{eqn:configl}
\end{equation}
where $\epsilon$ is a positive infinitesimal number.
Note that the mass formula is identical with (\ref{eqn:mass4})
when $(p,q)_n=(0,0)_1$,
and the charges of a particle corresponding to the string is $(0,0)_1$.
Thus we call the string and the particle $(0,0)_1$ in the following.

Now we solve the geodesics (\ref{eqn:configl}) numerically.
The resulting configurations are closed loops labeled by $z_*\geq 0$. 
As we decrease $z_*$, the loop gets smaller and finally collapses to a point
at $z=0$.
Note here that a loop of $z_*>1$ is closed not only in shape but
also in the string charges:
when the string crosses the cuts on $\mbox{Re}z>1$, the charges $(1,0)$
remain unchanged.
On the other hand, a loop of $0\leq z_*\leq 1$ is not closed in charges
because $(1,0)$ changes to $(1,-1)$ as it crosses the cut on 
$\mbox{Re}z\leq 1$ from the lower half plane.
Thus a string $(0,1)_1$ should be added to fill the difference of the charges.
From these facts we conclude that the configuration of $(0,0)_1$ is as 
depicted in Fig.5a when $z_*>1$,
Fig.5b when $0<z_*\leq 1$ and Fig.5c when $z_*=0$.
One can show that the strings in the figures have the same mass, $1/R$.
\begin{figure}
\hspace{0.5cm}
\epsfxsize=15cm
\epsfbox{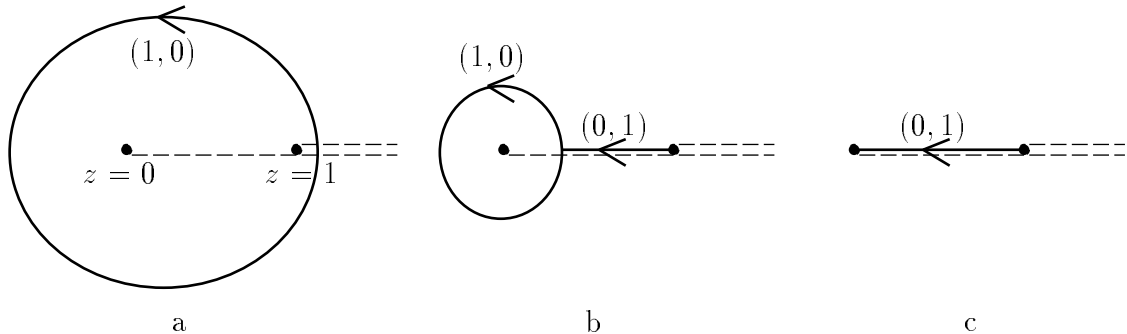}
\caption{Configurations of $(0,0)_1$.}
\end{figure}

From the figures we determine how $(0,0)_1$ is detected by the D3-brane probe
and find out the stability of the corresponding state.
First we focus on a part of $(0,0)_1$, a string $(1,0)$ which makes a loop. 
As we change $z_*$ from 0 to $\infty$, 
the string sweeps all the $z$-plane but $S$, the segment $[0,1]$ on the real 
axis.
Hence $(0,0)_1$ is detected through the string if the D3-brane is not on $S$.
The D3-brane on the loop does not divide $(0,0)_1$ in parts 
so that the string cannot decay. 
Therefore the corresponding state is stable in this case.
Next we consider the rest of $(0,0)_1$, the string $(0,1)_1$.
It appears only on $S$.
When we locate the D3-brane on it,
the brane separates $(0,0)_1$ into two parts, $(0,1)_1$ and $(0,-1)_0$.
Thus $(0,0)_1$ can decay to $(0,1)_1+(0,-1)_0$, and the corresponding state
is marginally stable.
In summary, the state $(0,0)_1$ exists for any value of the moduli
parameter $u$, 
and it is stable if $u$ is not on $S$ and marginally stable 
if $u$ is on $S$.

Next we shall generalize the results to the field theory with other 
parameters.
Recall that the parameters correspond to the positions of 7-branes.
Thus we deform the background (\ref{eqn:SWperiods}) by moving 7-branes,
and determine how the deformation affects the geodesics of loop strings.
Let us start with a large loop string in the background 
(\ref{eqn:SWperiods}) as depicted in Fig.5a.
In general, a geodesic of $(p,q)$ changes continuously unless it hits
a 7-brane $[r,s]$($\neq [p,q]$), while it changes to a string junction 
when it hits the 7-brane.
Therefore, if we move 7-branes finitely in the loop, 
the string remains a loop as depicted in Fig.6a.
Also if we drive a 7-brane $[1,0]$ to infinity, which corresponds
to reducing $N_f$, the loop of $(1,0)$ remains a loop.
On the other hand, if we drive $[2,-1]$ or $[0,1]$ to infinity, 
the loop string disappears:
the loop changes to a junction as depicted in Fig.6b
and the string connecting the loop and the 7-brane becomes infinitely long.
Thus the corresponding state has infinite mass, and disappears from
the spectrum.
This is natural because the deformations corresponds to the limit $R\rightarrow
0$.
In conclusion, a loop string large enough to encircle all the 7-branes 
exists in any background with finite $R$.
Next we fix the 7-branes and consider the region where the loop string sweeps. 
As we change the size of the loop from infinity to zero, the 
string hits 7-branes with different charges and changes to 
a junction as in Fig.6bc.
Then the singlet sweeps all the $z$-plane and finally it becomes 
$S$, the string junction with the vanishing loop as depicted in Fig.6d.
From the configurations we see that the singlet is stable when $u$ is 
not on $S$ and  marginally stable when $u$ is on $S$.
\begin{figure}
\hspace{0.5cm}
\epsfxsize=15cm
\epsfbox{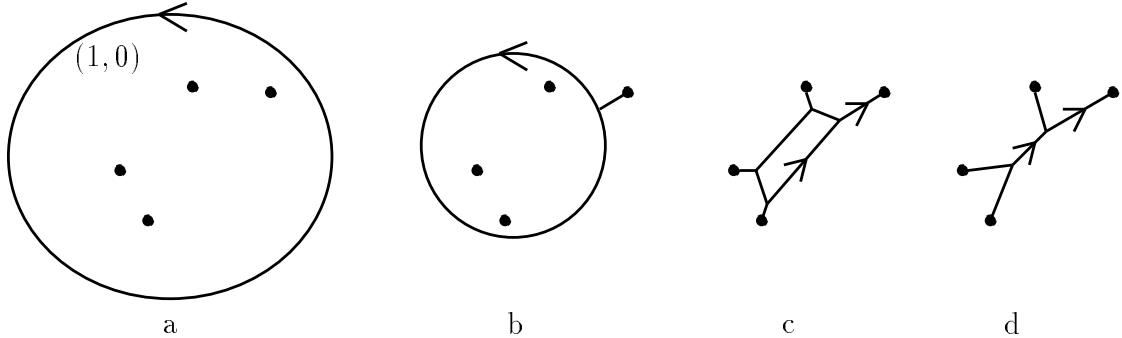}
\caption{Configurations of loop strings.}
\end{figure}

Although we do not discuss in detail, it is interesting to study the decays 
occurring on the D3-brane at the singularities.
As an example, let us consider the case where the D3-brane is 
located at $z=0$ in the background (\ref{eqn:SWperiods}).
Junctions in this background are constructed from $(0,1)_1$ 
and various $(p,q)_0$,
and the length of $(p,q)_0$ becomes zero when the D3-brane is located at $z=0$.
Thus the state $(P,Q)_n$ can decay to $(P,Q-n)_0+n(0,1)_1$.
Then $(0,1)_1$ can carry all the KK number of the theory.
Recall here that the configuration of $(0,0)_1$ observed at $z=0$ 
is the same with $(0,1)_1$ as depicted in Fig.5c.
The difference between $(0,1)_1$ and $(0,0)_1$ is that a massless 
string $(0,1)_0$ is added or not.
Thus the decay $(0,1)_1\leftrightarrow (0,0)_1+(0,1)_0$ would occur
and then all the KK number could be carried by the singlet.

\section*{Acknowledgements}

I would like to thank J.~Kamoshita, A.~Sugamoto, T.~Suzuki, T.~Uesugi 
and  H.~Yakazu for useful discussions. 
This work was partly supported by the Grant-in-Aid for Scientific Research on 
Priority Areas (No.14039204) from the Ministry of  Education, Culture, Sports,
Science and Technology, Japan.

\end{document}